\documentstyle[epsf,12pt]{article}
\begin{document}
\begin{center}
\Large{\bf Hadron Molecules Revisted}\\
\large{R.S. Longacre$^a$\\
$^a$Brookhaven National Laboratory, Upton, NY 11973, USA}
\end{center}
 
\begin{abstract}
Hadron Molecules are particles made out of hadrons that are held together by 
self interactions. In this report we discuss seven such molecules and their 
self interactions. The $f_0(980)$, $a_0(980)$, $f_1(1400)$, $\Delta N(2150)$ 
and $\pi_1(1400)$ molecular structure is given. We predict that two more states
the $K\overline{K}K(1500)$ and $a_1(1400)$ should be found.
\end{abstract}
 
\section{Introduction} 

The first two molecular states $f_0(980)$ and $a_0(980)$ are the isosinglet and
the isotriplet states of the $K$$\overline{K}$ bound system\cite{weinstein}. 
This binding requires a quark-spin hyperfine interaction in the over all 
$q$ $q$ $\overline{q}$ $\overline{q}$ system. We will see that this binding 
is different from the particle exchange mechanisms that bind the rest of 
molecules of this report. The exchanges of Ref.\cite{weinstein} that bind 
the $K$$\overline{K}$ system are quark exchanges where the quark-spin 
hyperfine interaction leads to an attractive potential. This attractive 
potential can only make states if the mesons of the fall apart mode 
$q$ $\overline{q}$ - $q$ $\overline{q}$ are below threshold. Thus we have 
states of $K$$\overline{K}$ lying just below threshold in scalar 
channel($0^{++}$). This is somewhat like the deuteron in the $p$ $n$ system.

The work of Ref.\cite{weinstein} has one flaw with regards to long-range color
van der Waals-type forces\cite{lipkin}. It has been pointed out that confining
potentials of the type used in Ref.\cite{weinstein} leads to a long-range 
power-law residual potential between color singlets of $r^{-2}$ between two 
mesons. Ref.\cite{weinstein} calculates the potential between mesons to be
given by
\begin{equation}
V_{vdW} = - {{20 MeV}\over{r^2}}.
\end{equation}
This is to be compared with the coulomb force between charged mesons
\begin{equation}
V_{coulomb} = - {{1.5 MeV}\over{r}}.
\end{equation}
The r in equation 1 and equation 2 is given in fm. Let us compare the potential
strength between them by setting them equal
\begin{equation}
V_{vdW} = V_{coulomb} = - {{20 MeV}\over{r^2}} = - {{1.5 MeV}\over{r}}.
\end{equation}
This occurs at the distance of 13.3 fm which is a very large distance. The
size of the scalar bound states are much smaller 2 fm at best. This long range
van der Waals force implies that gluons are massless(like the photon) and
can travel to the edge of the universe in a virtual state(like photons in 
EM-fields). Gluons can only try to travel to the edge if they are in a color 
singlet state (glueballs or glueloops). This would lead to an exponential 
cutoff
\begin{equation}
V_{r} = - {{e^{-\mu r}}\over{r}},
\end{equation}
where $\mu$ is the glueball mass. Sine mesons are much lighter than glueballs,
and the pion is the lightest, it is the longest carrier of the strong force. 
Meson exchanges will be the binding force of the other hadron molecules of this
report.

The report is organized in the following manner:

Sec. 1 is the introduction to $f_0(980)$ and $a_0(980)$. Sec. 2 looks at 
particle exchange calculations and the generation of $f_1(1400)$.  Sec. 3 
consider a dibaryon state $\Delta N(2150)$ and the similarity to $f_1(1400)$.
This similarity predicts the $a_1(1400)$ state. Sec. 4 an exotic meson state 
$1^{-+}$ which is seen in $\eta$ $\pi$ p-wave scattering $\pi_1(1400)$ is 
explained. Sec. 5 predicts another molecular state $K\overline{K}K(1500)$ 
which should be found. Sec. 6 is the summary and discussion.

\section{$K$$\overline{K}$$\pi$ as an interacting system} 

In order to bind $K$$\overline{K}$$\pi$ together, we need to develop a
dynamical theory that uses particle exchange mechanism not the quark exchange
that led to the van der Walls forces\cite{lipkin}. It is standard to break
up the scattering of a three particle system into a sum of isobar spectator 
scatterings\cite{herndon}. To complete this task we develop a unitary isobar
model which has long-range particle exchange forces. We will assume that the 
only interaction among the three particles are isobars decaying into two
particles where one of these particle exchange with the spectator forming
another isobar. This one-particle-exchange(OPE) occurs in the 
$K^*$$\overline{K}$, $\overline{K^*}$$K$ and $a_0$$\pi$ isobar systems.
See Figure 1(a) for the OPE mechanism(note $a_0$ is the $\delta$ isobar which
is its older name). We choose as our dynamical framework the 
Blankenbecler-Sugar formalism\cite{sugar} which yields set of coupled integral
equations for amplitudes $X(K^* \rightarrow K^*$), 
$X(K^* \rightarrow \overline{K^*}$), $X(K^* \rightarrow a_0$),
$X(\overline{K^*} \rightarrow K^*$), 
$X(\overline{K^*} \rightarrow \overline{K^*}$), 
$X(\overline{K^*} \rightarrow a_0$), $X(a_0 \rightarrow K^*$),
$X(a_0 \rightarrow \overline{K^*}$), and $X(a_0 \rightarrow a_0$).
These amplitudes describe the isobar quasi-two-body processes 
$K^* \overline{K} \rightarrow K^* \overline{K}$,
$K^* \overline{K} \rightarrow \overline{K^*} K$,
$K^* \overline{K} \rightarrow a_0 \pi$, etc., whose solution are Lorentz
invariant, and satisfy two- and three-body unitarity, and cluster properties.
In operator formalism these equations have the structure(a schematic 
representation is show in Figure 1(b)),  
\begin{equation}
X_{ba} = B_{ba}(W_E) + B_{bc}(W_E)G_c(W_E)X_{ca}(W_E) ;
a,b,c = K^*, \overline{K^*}, a_0.
\end{equation}
In equation 5, $W_E$ is the overall c.m. energy of the three-particle system.
The index $c$ which is summed over represent each isobar. Integration is over
solid angles where each total angular momentum is projected out as an 
individual set of coupled equations. The c.m. momentum of the spectator
which has the same magnitude as the isobar is integrated from zero to $\infty$.
Thus all effective masses of isobars are probed starting at the kinematic limit
down to negative $\infty$. All details are given in Ref.\cite{longacre}.
$G_c(W_E)$ is the propagator of isobar $c$ and $W_E$ determines the kinematic 
limit where c.m. momentum of the spectator and the isobar is zero. In Figure 2
we show the Breit-Wigner shape of the $K^*$ propagator $G_{K^*}(2.0)$ as a 
function of isobar mass($K \pi$). In Figure 3 we show the Breit-Wigner shape 
of the $a_0$ propagator $G_{a_0}(2.04)$ as a function of isobar 
mass($K \overline{K}$). 

\begin{figure}
\begin{center}
\mbox{
   \epsfysize 5.0in
   \epsfbox{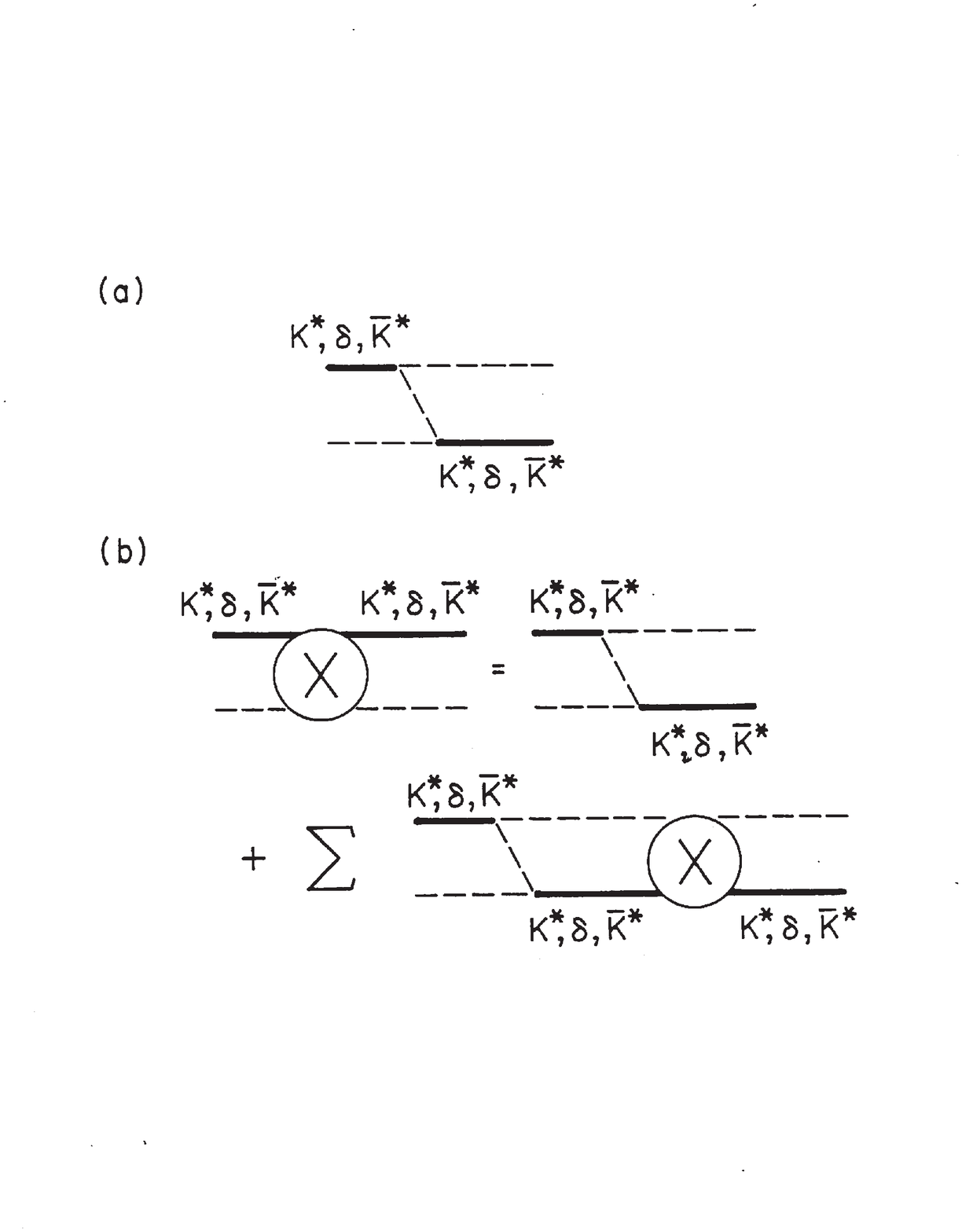}}
\end{center}
\vspace{2pt}
\caption{ (a) Long-range one-particle-exchange(OPE) mechanism. Isobars $K^*$,
$\overline{K^*}$, or $\delta$($a_0$ newer name) plus $\overline{K}$, $K$ or 
$\pi$ which absorbs the exchange particle which has decayed from the isobar
forming another isobar. (b) Unitary sum of OPE diagrams in terms of coupled 
integral equations.}
\label{fig1}
\end{figure}
 
\begin{figure}
\begin{center}
\mbox{
   \epsfysize 5.0in
   \epsfbox{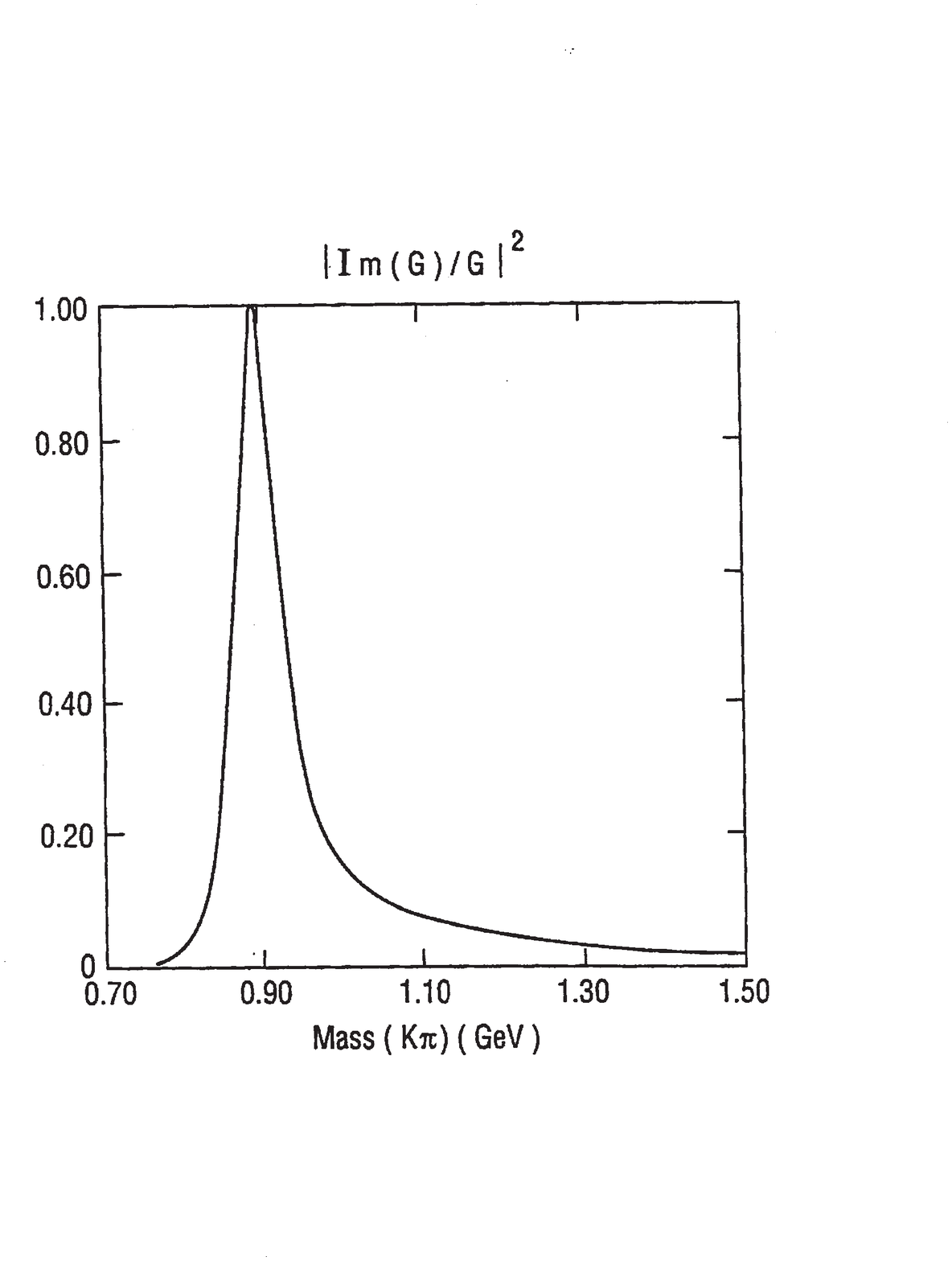}}
\end{center}
\vspace{2pt}
\caption{ The absolute value squared of the imaginary part divided by the
propagator for $K \pi$ propagation of the $K^*$ which is an 
$I = {{1}\over{2}}$ and $J = 1$ mode. This is equal to the square of the
T-matrix scattering of this $K \pi $ mode.}
 \label{fig2}
\end{figure}

\begin{figure}
\begin{center}
\mbox{
   \epsfysize 5.0in
   \epsfbox{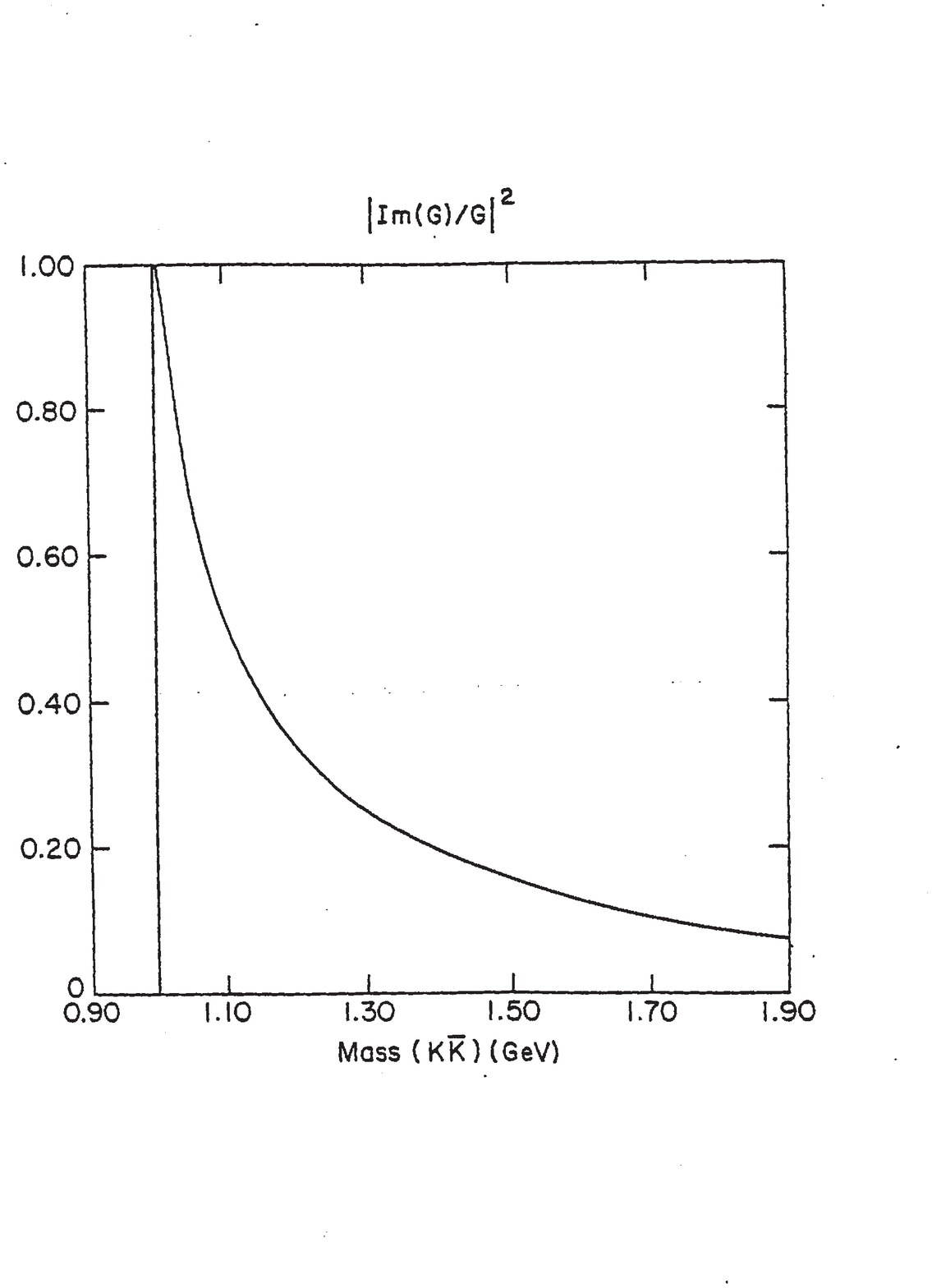}}
\end{center}
\vspace{2pt}
\caption{ The absolute value squared of the imaginary part divided by the
propagator for $K \overline{K}$ propagation of the $a_0$ which is an 
$I = 1$ and $J = 0$ mode. This is equal to the square of the
T-matrix scattering of this $K \overline{K} $ mode.}
\label{fig3}
\end{figure}

 Equation 5 can be rewriting as
\begin{equation}
\sum_k (\delta_{ik} - M_{ik}) X_{kj} = B_{ij} ;
i,j,k = K^*, \overline{K^*}, a_0.
\end{equation}
This Fredholm integral equation leads to a Fredholm determinant as a function
of $W_E$ for each partial wave or total $J$ projection. We have solved this
Fredholm determinant for two $J^{PC}$ states $0^{-+}$ and $1^{++}$. The results
of this analysis is shown in Figure 4. We see no binding effect in the 
$0^{-+}$ determinant, while in the $1^{++}$ channel there is a large effect
around $1.40$ GeV. At the energy of $1.40$ GeV the $K^*$ and $\overline{K^*}$
the peaks of Figure 2 are just coming into play. Since for $1^{++}$ they are 
in a s-wave they have maximum effect. In the $0^{-+}$ these peak are 
suppressed by a p-wave barrier. Around the mass $1.40$ GeV one can form a
picture of the system being $K$$\overline{K}$($a_0$) molecule at the center of
gravity with a light pion revolving in a p-wave orbit. The momentum of the pion
is such that at each half-revolution a $K^*$ or a $\overline{K^*}$ is 
formed(see Figure 5). The phase shift and production cross sections of this
molecular state is explored in detail in Ref.\cite{longacre}.
 
\begin{figure}
\begin{center}
\mbox{
   \epsfysize 5.0in
   \epsfbox{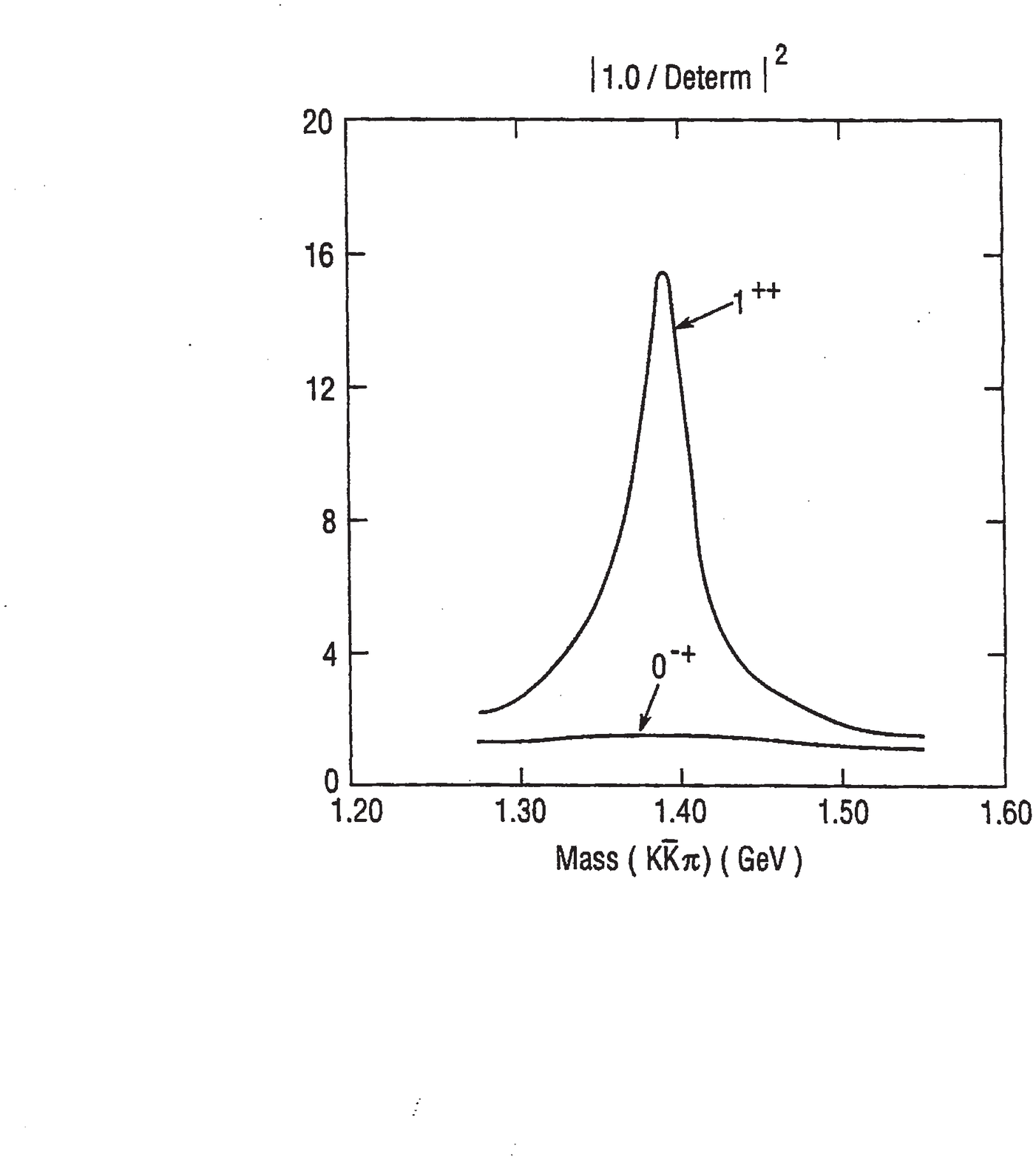}}
\end{center}
\vspace{2pt}
\caption{The value of 1 over the Fredholm determinant squared for 
$J^{PC}$ = $1^{++}$ and $J^{PC}$ = $0^{-+}$  as a function of 
$K$$\overline{K}$$\pi$ mass(smooth curves).}
\label{fig4}
\end{figure}

\begin{figure}
\begin{center}
\mbox{
   \epsfysize 6.0in
   \epsfbox{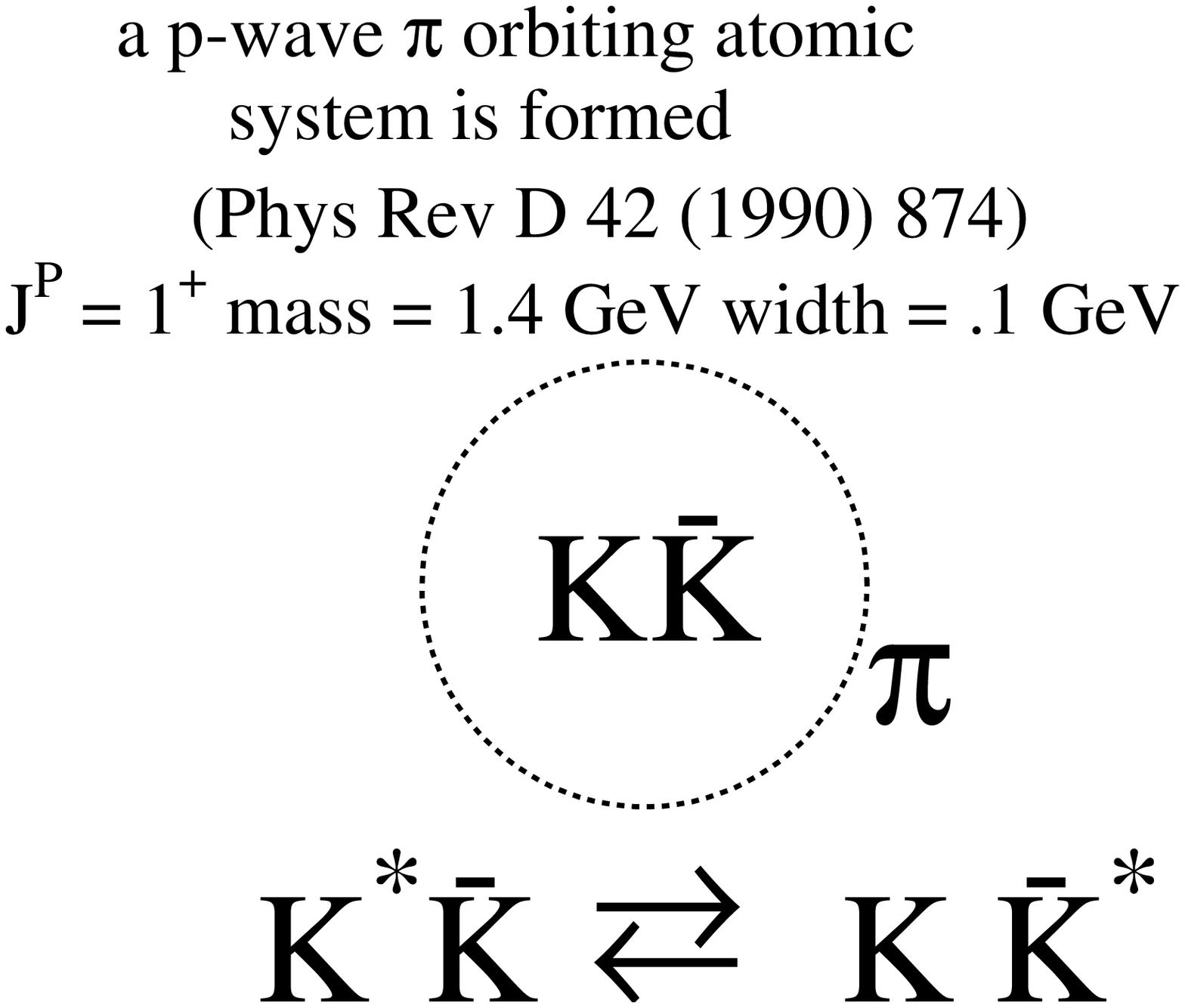}}
\end{center}
\vspace{2pt}
\caption{The meson system mainly resonates in the s-wave $K^*$$\overline{K}$ 
and $K$$\overline{K^*}$ mode with a pion rotating in a p-wave about a 
$K$ $\overline{K}$ system which forms a isospin triplet. The pion moves back 
and forth forming $K^*$ and $\overline{K^*}$ states with one $K$ or 
$\overline{K}$.}
\label{fig5}
\end{figure}

\begin{figure}
\begin{center}
\mbox{
   \epsfysize 6.0in
   \epsfbox{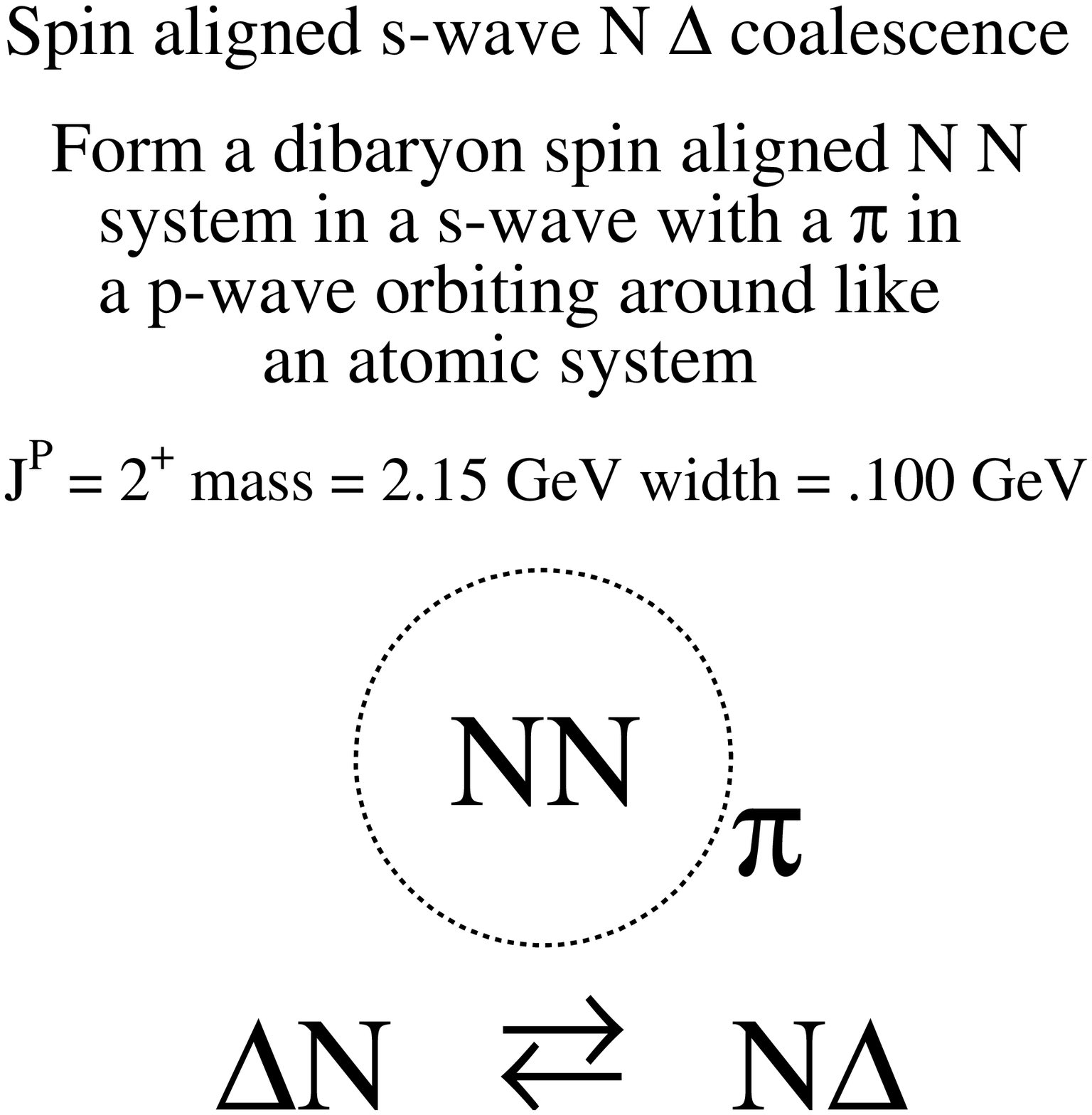}}
\end{center}
\vspace{2pt}
\caption{The dibaryon system mainly resonates in the s-wave $\Delta$ $N$ mode 
with a pion rotating in a p-wave about a spin aligned $N$ $N$ system which 
forms a isospin singlet. The pion moves back and forth forming $\Delta$ states 
with one nucleon and then the other.}
\label{fig6}
\end{figure}

\section{Two more molecular states}

\subsection{Dibaryon state $\Delta N(2150)$ as a molecule.}

The dibaryon state interacts in three two-body scattering channels. Its mass 
is 2.15 GeV and has a strong interaction resonance decay width of 100 MeV. It 
interacts in the $N$$N$ d-wave spin anti-aligned\cite{Arndt}, $d$$\pi$ p-wave
spin aligned\cite{Oh}, and $\Delta$$N$ s-wave spin aligned\cite{Schiff}. The 
dibaryon system mainly resonates in the s-wave $\Delta$$N$ mode with a pion
rotating in a p-wave about a spin aligned $N$$N$ system which forms a isospin
singlet. The pion moves back and forth forming $\Delta$ states with one nucleon
and then the other(see Figure 6). All three isospin states of the pion can be 
achieved in this resonance. Thus we can have $\pi^+$$d$, $\pi^0$$d$, and 
$\pi^-$$d$ states. If the pion is absorbed by any of the nucleons it under 
goes a spin flip producing a d-wave $N$$N$ system. The resonance decays into 
$N$$N$, $\pi$$d$, or $\pi$$N$$N$. In the last section we saw a meson system
that had an analogous orbiting pion in a p-wave mode about a 
$K \overline{K}$ in a s-wave\cite{longacre}. Both systems have 
a similar lifetime or width of $\sim$ .100 GeV\cite{deuteron}.  

\subsection{$a_1(1400)$ state is predicted}

Unlike the $f_1(1400)$ the $\Delta N(2150)$ has an isosinglet at the center
of motion. The $K$$\overline{K}$ isosinglet state of Sec. 1 could form the
center of motion for an isotriplet molecular state $a_1(1400)$. The set of 
integral equation would be the same as in the $f_1(1400)$ case making a 
similar Fredholm determinant. Like for $\Delta N(2150)$ which had a $d \pi$
decay mode, one would expect that there would be a $f_0(980)$ $\pi$ decay
mode. We can calculate the branching ration of $f_1(1400)$ to $a_0$$\pi$
from the Dalitz plot calculated using equation 20 of Ref.\cite{longacre}.
The ratio in the plot going into $a_0$$\pi$ is 22\%. The reason this mode
is so small is because $\sqrt{Imag(D_{a_0})}\over{|D_{a_0}|}$ is much
smaller than $\sqrt{Imag(D_{K^*})}\over{|D_{K^*}|}$\cite{longacre}. 
Where as the ratios of $Imag(D_{a_0})\over{|D_{a_0}|}$ and 
$Imag(D_{K^*})\over{|D_{K^*}|}$ are one at resonance(see Figure 2 and 3). 
For the $\Delta N(2150)$ the $d \pi$ branching ratio is 25\%\cite{deuteron}.
We should expect that the branching of $a_1(1400) \rightarrow f_0 \pi$ should 
be the same as $f_1(1400) \rightarrow a_0 \pi$. Dr. Suh-Urk Chung has 
claimed such a state has been observed\cite{suh-urk}.

\section{Exotic state $J^{PC}$ = $1^{-+}$ $\pi_1(1400)$ as a molecule.}

In Sec. 2 we explained the $f_1$(1420) seen in 
$\overline{K} K\pi$\cite{longacre}. Following the same approach we can 
demonstrate the possibility that the $\pi_1$(1400) is a $\overline{K} K\pi\pi$ 
molecule, where the $\overline{K} K\pi$ in a relative s-wave with the other 
$\pi$ orbiting them in a p-wave. Since the $\overline{K} K\pi$ is 
resonating as the $\eta$(1295), it is possible that the offshell $\overline{K}
K\pi(\eta)$ would couple to the ground state $\eta$, thus creating a $\eta\pi$ 
p-wave decay mode. 

As was done in Ref.\cite{longacre}, we need to arrange a set of Born terms 
connecting all of the possible intermediate isobar states of the 
$\overline{K} K\pi\pi$ system ($\eta$(1295)$\pi$, $a_0$(980)$\rho$(770), 
$K_1$(1270)$\overline{K}$ or $\overline{K_1}(1270)$$K$). We assume that 
the only interaction among the particles occurs through one-particle exchange 
(OPE), thus connecting the above isobar states (Figure 7). In order to 
completely derive the dynamics one would have to
develop a true four-body scattering mechanism with OPE Born terms connecting
two- and three-body isobar states. We can take a short cut and use the
three-body formalism developed in Ref.~\cite{longacre}, if we note that the set
of diagrams (Figure 8) could be summed using a true four-body formalism, and 
be replaced by the Born term of Figure 9. Here the $a_0$(980) is treated as a 
stable particle and the $\pi\pi$ p-wave phase shift ($\rho_{med}$) is 
assumed to be modified by the sum of terms in Figure 8. With this assumption, 
then binding can occur if we use the $N/D$ propagators for the 
$\eta$(1295)(see Figure 10) and $\rho_{med}$(see Figure 11). In Figure 11
we also show the unaltered p-wave phase shift ($\rho$). Figure 12 shows the 
final state enhancement times the $\eta$(1295) $\pi$ p-wave kinematics. 
The bump is driven by the collision on the Dalitz plot of the $\eta$(1295) 
Breit-Wigner (Figure 10) and the rapid increase of the $\pi\pi$ p-wave phase 
shift (Figure 11).

We have suggested the possibility that the $\pi_1$(1400) is a final state    
interaction for the $K\bar K\pi$ system in a s-wave orbiting by a $\pi$ in 
a p-wave. The $\eta\pi$ decay mode is generated by the off shell appearance 
of the $\eta$ from the $K\bar K\pi$ system ($0^{-+}$). Our model thus 
predicts that a strong $J^{PC}=1^{-+}$ should be seen in the $K\bar K\pi\pi$ 
system at around 1.4 GeV/c$^2$. If the $\pi_1$(1400) is only seen in the 
$\eta\pi$ channel then its hard to understand three facts about its 
production. First, that the force between the $\eta$ and $\pi$ in a p-wave 
should be repulsive (QCD)~\cite{barnes}. This is not a problem if the 
$\eta\pi$ is a minor decay mode. Second, why should the production be so small 
compared to the $a_2$  which has only a 14\% branching to $\eta\pi$? One would 
think it should be produced in unnatural parity exchange not natural. Again 
this is not a problem if minor decay mode. 

\begin{figure}
\begin{center}
\mbox{
   \epsfysize 6.0in
   \epsfbox{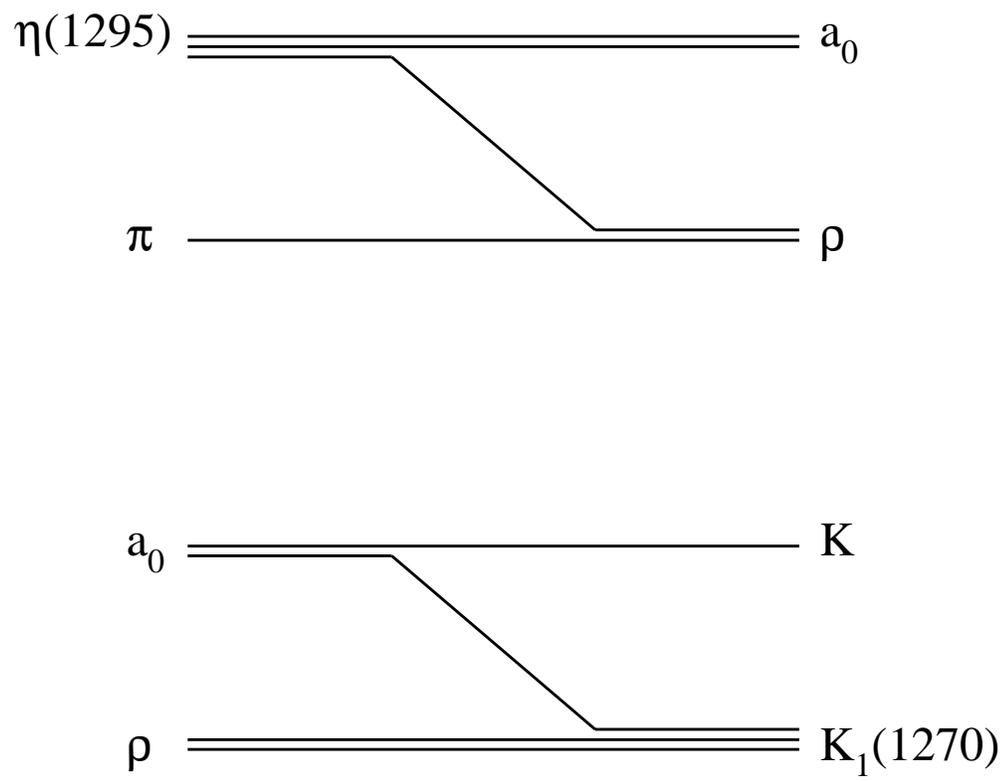}}
\end{center}
\vspace{2pt}
\caption{One-particle-exchange (OPE) Born terms for 
$\overline{K} K\pi\pi$ system.}
\label{fig7}
\end{figure}

\begin{figure}
\begin{center}
\mbox{
   \epsfysize 6.0in
   \epsfbox{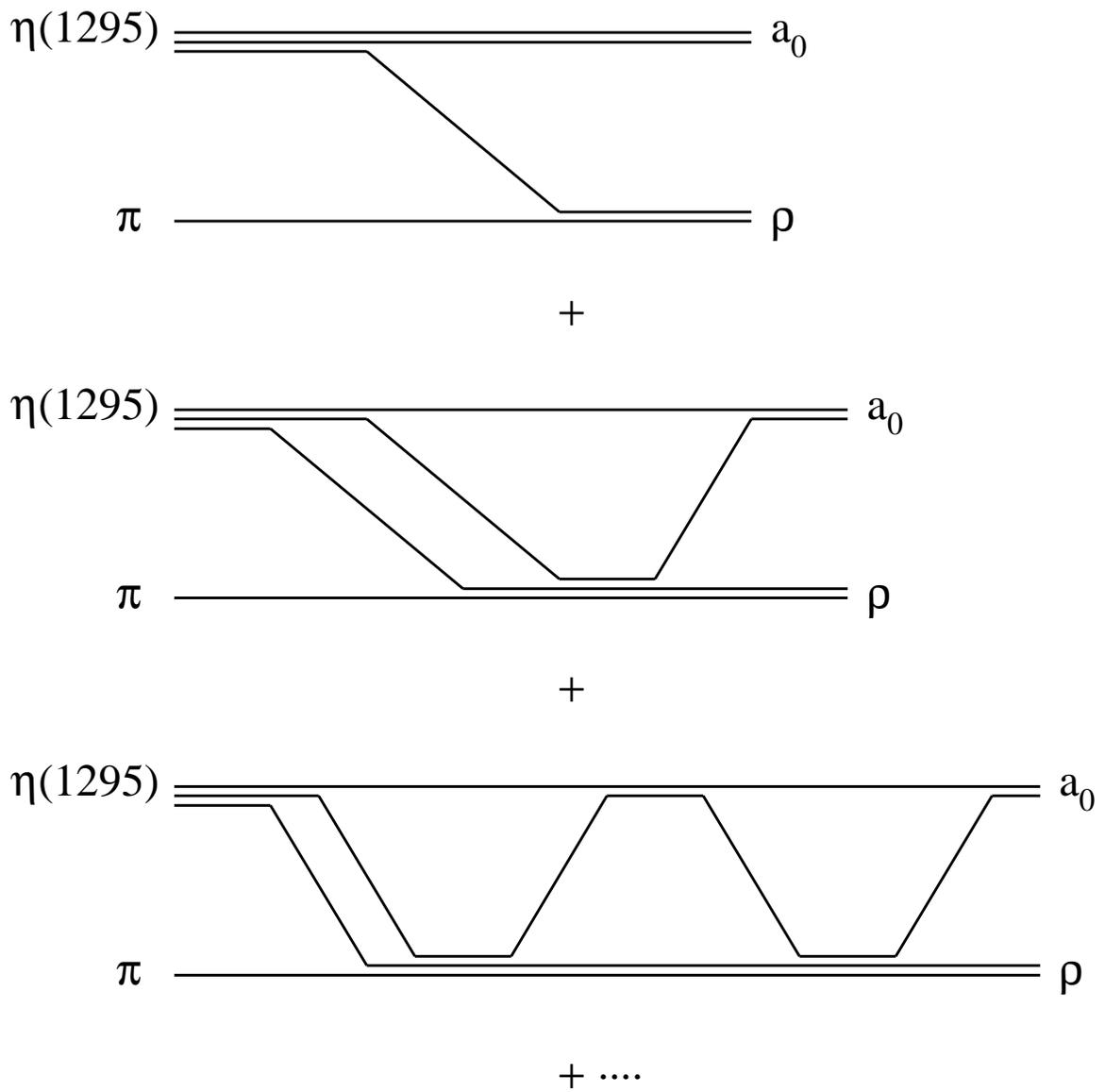}}
\end{center}
\vspace{2pt}
\caption{The set of infinite terms where all $K$ and $\overline{K}$ exchanges 
are summed.}
\label{fig8}
\end{figure}

\begin{figure}
\begin{center}
\mbox{
   \epsfysize 7.0in
   \epsfbox{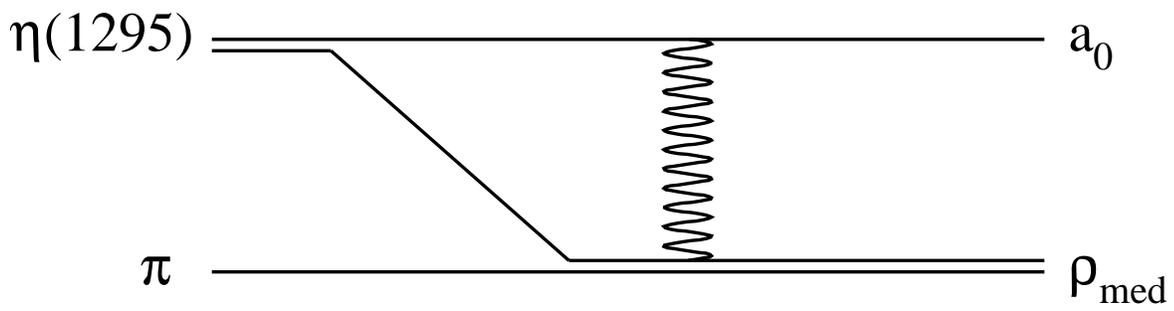}}
\end{center}
\vspace{2pt}
\caption{The Born that is used in the three-body effective analysis, where the 
$\pi\pi$ p-wave is altered by the sum of terms in Figure 8.}
\label{fig9}
\end{figure}

\begin{figure}
\begin{center}
\mbox{
   \epsfysize 6.0in
   \epsfbox{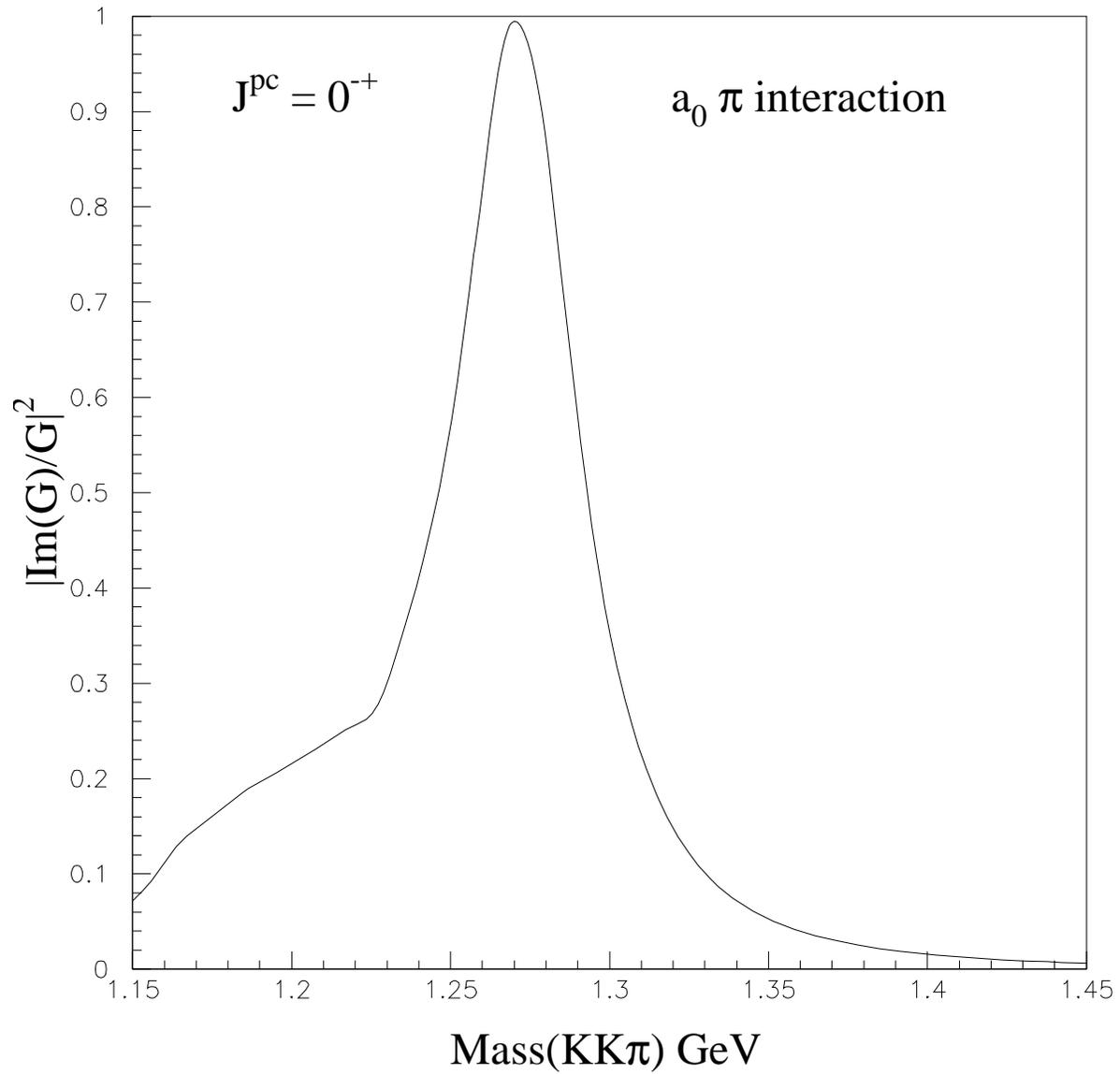}}
\end{center}
\vspace{2pt}
\caption{The absolute value squared of the imaginary part of the 
$\eta$ (1290) propagator divided by the complete propagator, thus forming the 
square of the T-matrix scattering amplitude.}
\label{fig10}
\end{figure}

\begin{figure}
\begin{center}
\mbox{
   \epsfysize 6.0in
   \epsfbox{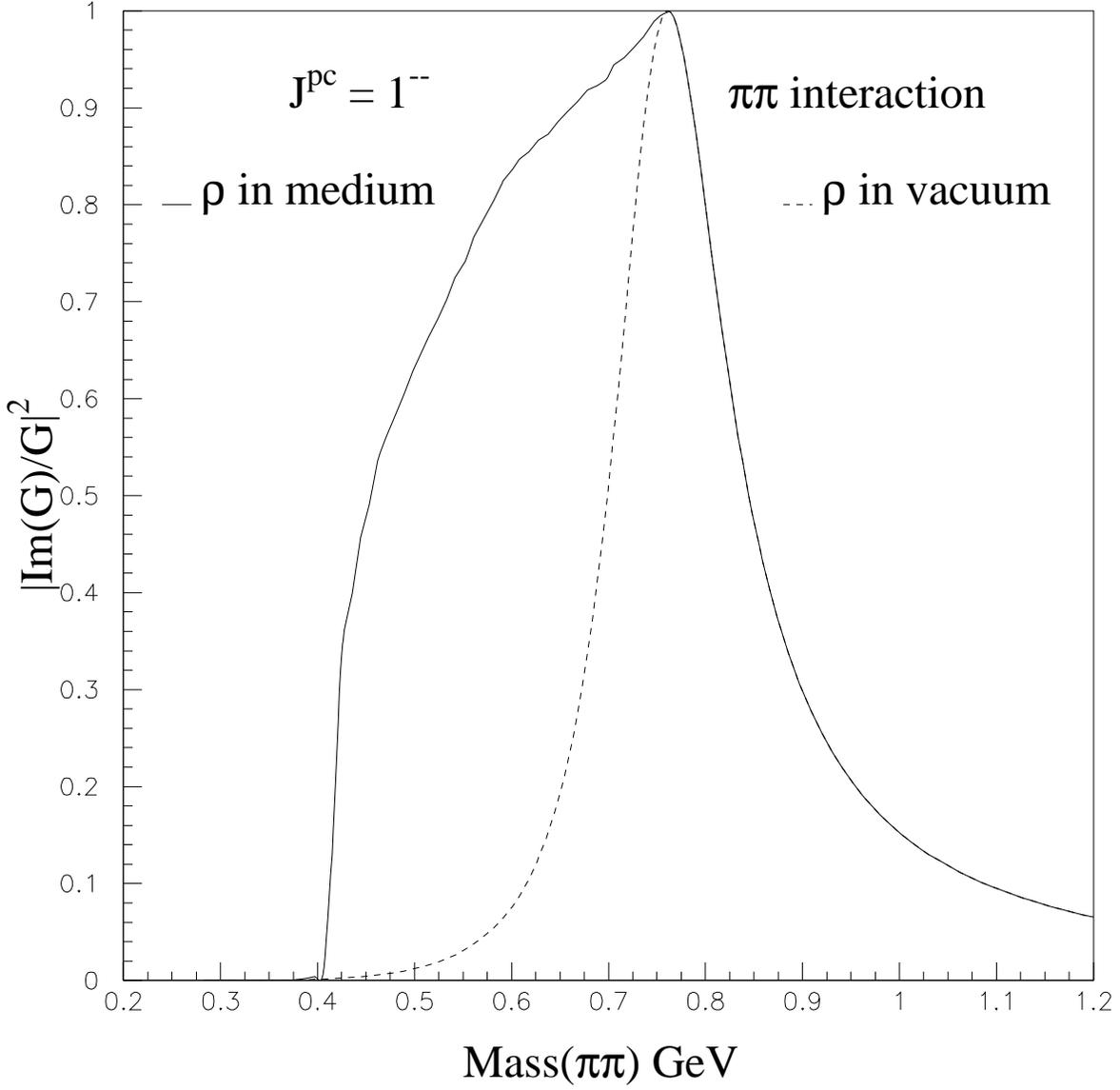}}
\end{center}
\vspace{2pt}
\caption{The absolute value squared of the imaginary part of the $\pi\pi$ 
p-wave phase shift: the solid line is the modified phase shift; the dashed 
line is the original vacuum phase shift which is the $\rho$ meson.}
\label{fig11}
\end{figure}

\begin{figure}
\begin{center}
\mbox{
   \epsfysize 6.0in
   \epsfbox{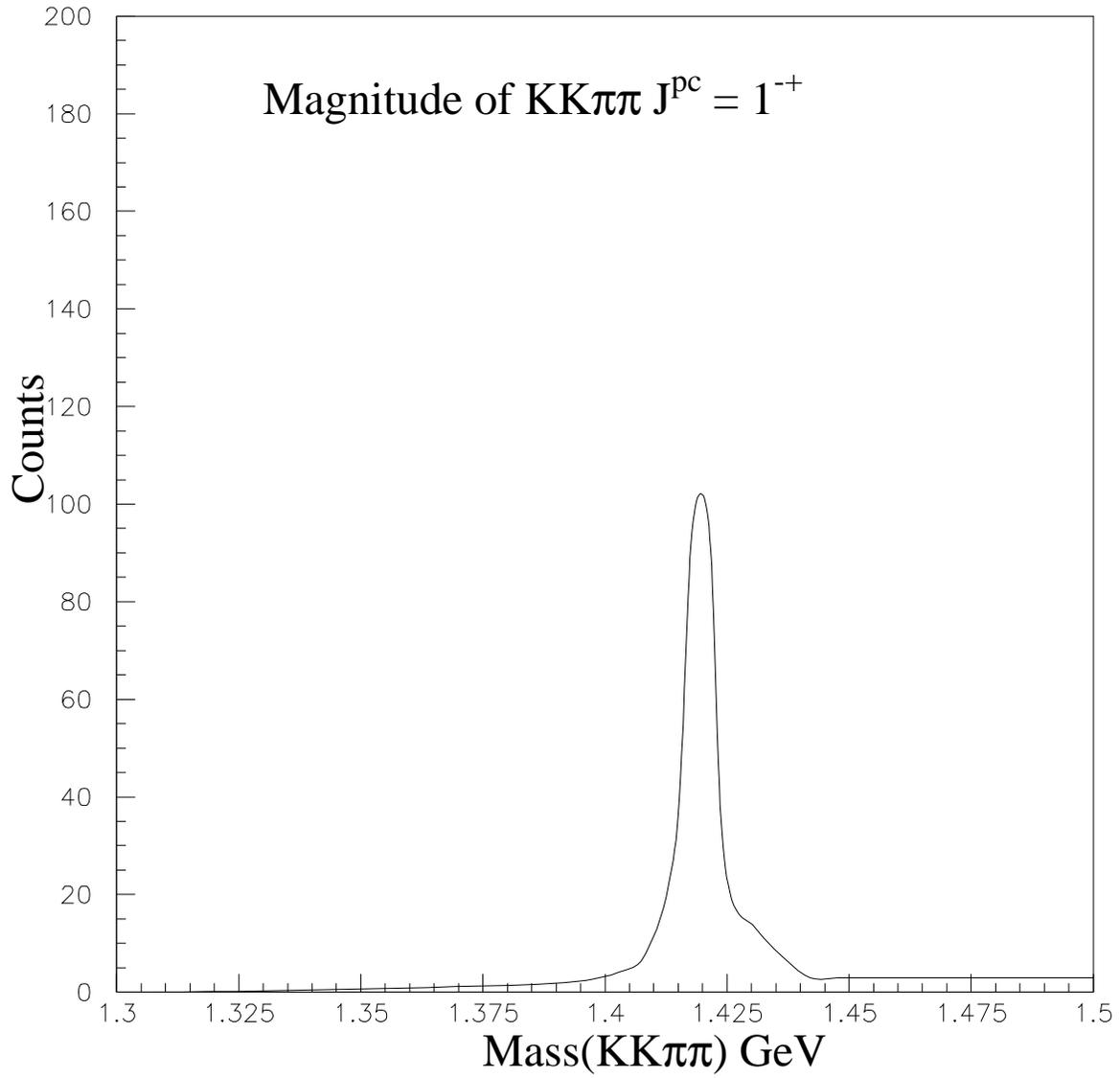}}
\end{center}
\vspace{2pt}
\caption{The value of 1 over the Fredholm determinate squared times the
kinematics of p-wave $\pi\eta$(1290).}
\label{fig12}
\end{figure}

\clearpage

Finally, it is reasonable to think that the largest decay amplitude would be 
the modes that have an $a_0$(980) in the final state. However in 
Ref.\cite{longacre} the same conclusion was initially drawn, except when one 
puts in all the numerical factors the $a_0$  modes become suppressed.  The 
explanation comes from the very powerful attraction of the kaons in the $a_0$  
mode. The isobar decay amplitude is proportional to $\sqrt{N}/D$ both $N$ and 
$D$ are large numbers while the ratio is near one at the threshold(see last 
section and Ref.\cite{longacre}). Thus the decay amplitude becomes 
proportional to $1/\sqrt{N}$ . We predict that the major mode could be 
$\pi \pi$ p-wave having no $\rho$ peak (work above) forming a $K\pi\pi$ or 
a $\overline{K}\pi\pi~J^p = 1^+$  plus a $\overline{K}$ or $K$ 
with overall $G$-parity minus. The $K\pi\pi$ should more or less be a phase 
space distribution.

\section{$K\overline{K}K(1500)$ state is predicted}

We saw in Sec.1 and Sec. 2 that the $K$$\overline{K}$ system had attraction
through the $a_0(980)$ resonance. It seems only natural to investigate the 
possibility that a three-K molecule might exist. This is only worthwhile
if we consider only exotic quantum numbers. The only exotic quantum number 
which can be be obtained is the isotopic spin. Thus a set of coupled 
equations for the $K$$\overline{K}$$K$ system in a overall s-wave with 
isotopic spin of $3\over{2}$ is created\cite{longacre}. The Fredholm 
determinate squared times of the equations is shown in Figure 13.

Isopin spin $3\over{2}$ implies there are four states 
$K^+$$\overline{K^0}$$K^+$,$K^+$$K^-$$K^+$,
$K^0$$\overline{K^0}$$K^0$, and $K^0$$K^-$$K^0$. The 
$K^+$$\overline{K^0}$$K^+$ is double charged. There would also be a 
$K^-$$K^0$$K^-$ which is the anti-matter state of the 
$K^+$$\overline{K^0}$$K^+$. These states are unique to this type of
binding mechanism.

\section{Summary and Discussion}

In this report we have discussed seven possible hadron molecular states. These
states are particles made out of hadrons that are held together by 
self interactions. The seven molecules and their self interactions are 
explored. The $f_0(980)$, $a_0(980)$ relied quark exchange forces which
made states of $K$$\overline{K}$ lying just below threshold in scalar 
channel($0^{++}$). This is somewhat like the deuteron in the $p$ $n$ system.
The $f_1(1400)$, $\Delta N(2150)$ and $\pi_1(1400)$ molecular structure are
held together by long range particle exchange mechanisms not the quark 
exchange that led to the van der Walls forces\cite{lipkin}. These exchange
mechanisms also predicts that two more states the $K\overline{K}K(1500)$ and 
$a_1(1400)$ should be found. For the $a_1(1400)$ the set of integral equation 
would be the same as in the $f_1(1400)$ case making a similar Fredholm 
determinant. Like for $\Delta N(2150)$ which had a $d \pi$
decay mode, one would expect that there would be a $f_0(980)$ $\pi$ decay
mode. Dr. Suh-Urk Chung has claimed such a state has been 
observed\cite{suh-urk}.

\begin{figure}
\begin{center}
\mbox{
   \epsfysize 6.13in
   \epsfbox{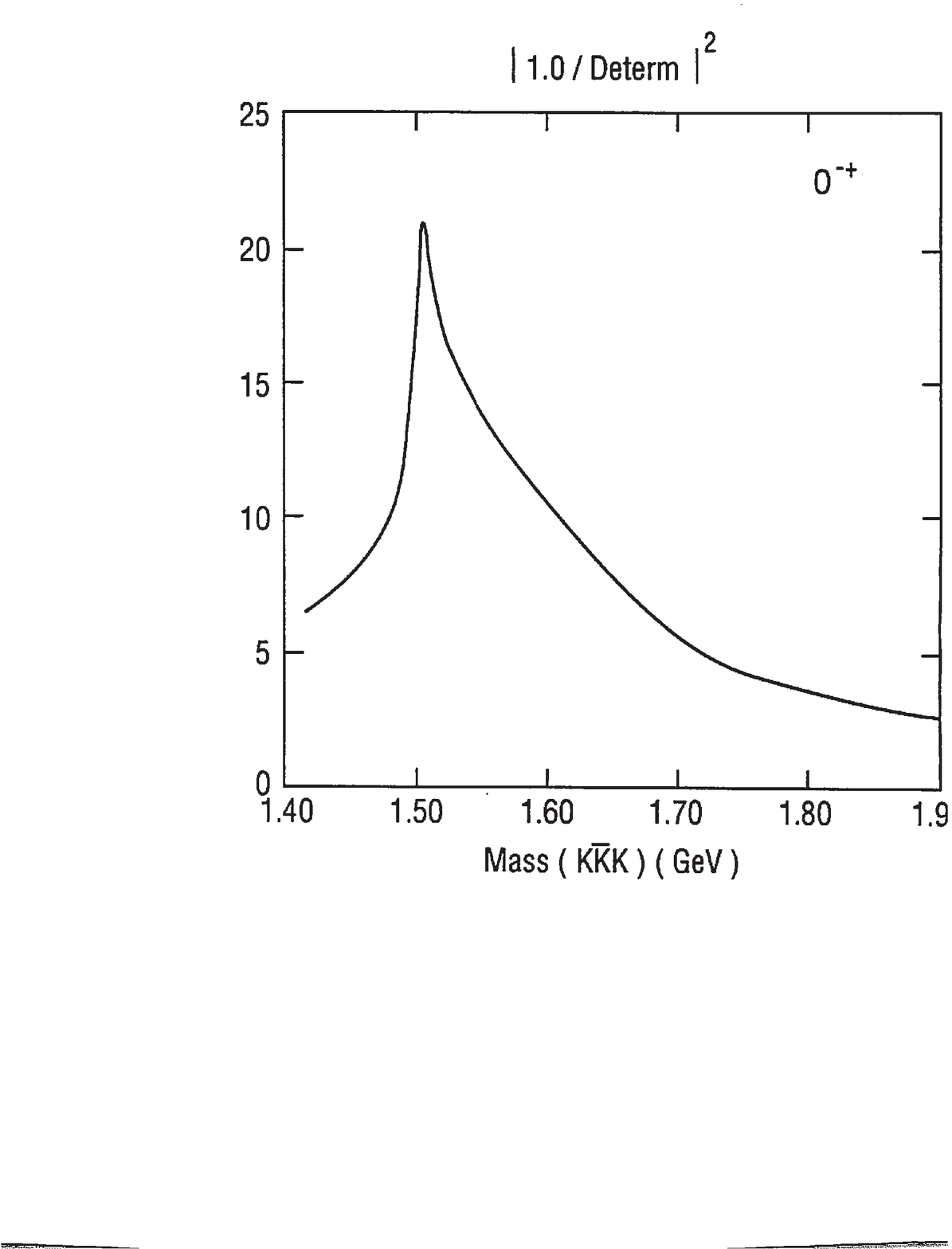}}
\end{center}
\vspace{2pt}
\caption{The value of 1 over the Fredholm determinant squared for 
$J^{P}$ = $0^-$  as a function of $K$$\overline{K}$$K$ mass(smooth curves).}
\label{fig13}
\end{figure}

\section{Acknowledgments}

This research was supported by the U.S. Department of Energy under Contract No.
DE-AC02-98CH10886.

\end{document}